\documentclass[twocolumn,prb]{revtex4-1}
\usepackage[latin9]{inputenc}
\usepackage{amsmath}
\usepackage{amssymb}
\usepackage{graphicx}
\usepackage{esint}
\usepackage[unicode=true,pdfusetitle,
 bookmarks=true,bookmarksnumbered=false,bookmarksopen=false,
 breaklinks=false,pdfborder={0 0 1},backref=false,colorlinks=false]
 {hyperref}

\makeatletter

\providecommand{\tabularnewline}{\\}
 \@ifundefined{textcolor}{}
 {%
   \definecolor{BLACK}{gray}{0}
   \definecolor{WHITE}{gray}{1}
   \definecolor{RED}{rgb}{1,0,0}
   \definecolor{GREEN}{rgb}{0,1,0}
   \definecolor{BLUE}{rgb}{0,0,1}
   \definecolor{CYAN}{cmyk}{1,0,0,0}
   \definecolor{MAGENTA}{cmyk}{0,1,0,0}
   \definecolor{YELLOW}{cmyk}{0,0,1,0}
 }

\makeatother

\begin{document}

\title{Slow non-exponential phase relaxation and enhanced mesoscopic kinetic
inductance noise in disordered superconductors}

\author{K. Kechedzhi}

\affiliation{$^{1}$ Condensed Matter Theory Center and Joint Quantum Institute,
Department of Physics, University of Maryland, College Park, 20742
MD, USA}
\begin{abstract}
Mesoscopic low frequency noise in electrical characteristics of disordered
conductors is a result of dynamic quantum interference pattern due
to motion of defects. This has been firmly established by demonstrating
the characteristic partial suppression of the noise amplitude by the
dephasing effect of a weak external magnetic field. The spatial correlation
of the quantum interference pattern in disordered normal state conductors
is invariably limited by the exponential phase relaxation due to inelastic
processes. In this paper we develop a quantitative theory of the mesoscopic
noise in the s-wave superconducting phase of a strongly disordered
superconductor (such that the superconducting coherence length is
much longer than the mean free path). We find that the superconducting
coherence length limits the quantum interference effects in superconductors.
However, in contrast to the normal phase, the decay of the phase relaxation
on the scale of the superconducting coherence length is non-exponential.
This unusual slow relaxation manifests in the enhanced amplitude of
the mesoscopic noise in superconductors and a peculiar non-linear
scaling of the amplitude with the strength/number of mobile defects
in very thin superconducting films and wires (effectively 2D and 1D
with respect to the superconducting coherence length). Mesoscopic
noise sets a natural limit on the quality of kinetic inductance elements. 
\end{abstract}
\maketitle

\section{Introduction}

Development of superconducting circuits with very low levels of noise
is largely motivated by their potential applications for quantum computing
devices\cite{Nori2005,Clarke2008}, ultra-sensitive detectors\cite{Zmuidzinas2004,100years}
and magnetometers\cite{ClarkeBook}. Performance of all these devices
is limited by the level of intrinsic noise of various types. A dramatic
progress was achieved recently with the elimination of the amorphous
insulators and/or suppression of the charge noise associated with
them\cite{Martinis2005,Schreier2008,Manucharyan2009}. This requires
minimizing the use of arrays of Josephson junctions as inductance
elements in these circuits. The use of magnetic self-inductance of
wires and coils is not feasible due to size and geometry restrictions
in these devices. High kinetic inductance appears naturally in disordered
superconductors which makes them an obvious candidate for these elements.
Here we show that disordered superconducting wires show significant
fluctuations of kinetic inductance due to electron interference induced
by defect motion; we develop a quantitative analytical theory of this
quantum interference effect and calculate the amplitude of the kinetic
inductance fluctuations and noise with accurate numerical coefficients.
The noise due to these fluctuations provides a natural limit for the
quality of kinetic inductance elements\cite{WallraffSchoelkopf2004,Klapwijk2013,AnnunziataCatelani2010,Zmuidzinas2012}.

Kinetic inductance noise is caused by local fluctuations of impurities
invariably present in metallic wires. Quantum interference of electrons
moving diffusively in the potential of impurities significantly enhances
the effect of these local fluctuations. This is because the macroscopic
interference pattern in a conductor is very sensitive to the position
of individual impurities. Motion of a single impurity, for example
an impurity jumping between two stable spatial configurations, results
in a substantial fluctuation in macroscopic (or rather mesoscopic)
properties of the conductor. This effect was analyzed in great detail
for the case of metals in the normal state\cite{AltshulerSpivak,Feng86}
in which the quantum interference leads to mesoscopic noise, i.e.
a substantial enhancement of the noise in electronic characteristics
due to local fluctuations in the impurity potential. The quantum interference
pattern in the normal state is spatially correlated up to the length
scale, $L_{\varphi}$, that limits the coherent propagation of electrons.
$L_{\varphi}$ is typically set by low energy electron-electron or
electron-phonon scattering\cite{LeeStoneFukuyama,AAKH}. In superconductors,
where electrons at low energies are bound into Cooper pairs similarly
to the case of normal metals macroscopic characteristics demonstrate
mesoscopic fluctuations\cite{SpivakZyuzin,Koshelev2012}. However,
the spatial correlations of mesoscopic fluctuations in a superconductor
are distinct from those in normal metals. Paired state of electrons
in presence of disorder is characterized by a superconducting coherence
length $\zeta\equiv\sqrt{D(1/\Delta)}$ where $D$ is the diffusion
coefficient and $\Delta$ is the superconducting gap, which describes
a typical diffusion path of an electron during a time $1/\Delta$,
a lifetime of a virtual excitation with the energy of order $\Delta$.
It is reasonable to expect that the role of the dephasing length in
superconductors is played by the superconducting correlation length,
i.e. $\zeta$ defines the scale of exponential phase relaxation and
limits the quantum interference responsible for the variation of macroscopic
characteristics. However, we will show that this intuition is not
fully consistent with microscopic calculations. Instead, we find that
the superconducting gap $\Delta$ limits the coherence of electronic
states however the phase relaxation has a slow non-exponential character.
The slow phase relaxation results in an enhanced amplitude of noise
in superconductors and non-linear scaling of the noise amplitude with
the strength/number of locally fluctuating impurities in thin films
and wires. This behavior is in contrast to the exponential phase relaxation
in the case of weak localization corrections to the superfluid density
of s-wave superconductors\cite{SmithAmbegaokar}.

Experimentally, the noise in the normal state conductivity associated
with quantum interference is a well established phenomena\cite{Birge1989,Birge1990,Birge1993}.
In contrast, the noise in kinetic inductance of Josephson circuits
was reported to be absent in early experiments\cite{Wellstood1987}
and observed only very recently in Ref.~\onlinecite{McDermott}.
This work\cite{McDermott} also reports a surprising degree of correlations
between the inductance noise and flux noise in the superconducting
devices that remain poorly understood\cite{Kechedzhi_2011}.

The main result of this paper is the quantitative prediction of the
average amplitude $\delta_{K}^{2}\equiv\left\langle (K_{u'}-K_{u})^{2}\right\rangle /(2\left\langle K\right\rangle ^{2})$
of the mesoscopic noise in the kinetic inductance $K$ in superconducting
wires. Here $K_{u}$ and $K_{u'}$ correspond to the value of kinetic
inductance given impurity potential $u$ and $u'$ respectively. In
realistic measurement the two distinct impurity potential realization
$u$ and $u'$ represent different moments in time, and the amplitude
$\delta_{K}^{2}$ is averaged over long periods of time, longer than
any characteristic time of the electronic system, and longer than
the characteristic fluctuation time of impurities. Practically, this
amplitude can be connected to the amplitude of the noise spectral
density.\footnote{Here we assume that the observation time $t\rightarrow \infty$ is long enough. The noise amplitude $\delta_K^2$ can be connected to the noise power spectra typically measured experimentally. We assume that scattering centers responsible for the noise can be modelled by a system of bistable tunneling defects with a Lorentzian noise spectrum characterized by a tunneling time $\tau$ with the distribution $g(\tau)$,
$\mathcal{P}_K(\omega)=\delta_K^2\int d\tau g(\tau)\frac{2\tau}{1+(\omega\tau)^2}$. For a typically considered model of $g(\tau)=A/\tau$ for $\tau_{min}<\tau<\tau_{max}$ and the normalization coefficient given by $A=1/(\pi\ln(\tau_{max}/\tau_{min}))$. For this model the amplitude of the noise can be related to the power spectra,
$
\mathcal{P}_K(\omega)=\delta_K^2\frac{1}{\ln(\tau_{max}/\tau_{min})} \frac{1}{\omega}.
$
This defines the meaning of $\gamma_T$ which is the density of defects integrated over a broad bandwidth.} 

We consider a finite size rectangular piece of a superconductor of
a mesoscopic size $L_{x}\times L_{y}\times L_{z}$. In other words
the dimensions of the superconductor are not too much larger than
the superconducting coherence length. We show that in three dimensional
samples, $\zeta\ll L_{x},\, L_{y},\, L_{z}$ the result is given by,

\begin{gather}
\delta_{K}^{2}\approx\widetilde{C}_{3}\gamma_{T}\frac{1}{(p_{F}\ell)^{4}}\frac{\zeta^{3}}{V},\label{3Dnoise}
\end{gather}
where $p_{F}$, is Fermi wave vector, $\ell$ is the mean free path,
$\zeta=\sqrt{D/\Delta}$ coherence length of the superconducting electrons,
$V=L_{x}L_{y}L_{z}$ is the wire volume and the numerical coefficient
is $\widetilde{C}_{3}\approx60$. The relative density of thermally
activated defects is defined by $\gamma_{T}=\Gamma_{T}/\Gamma$, where
$\Gamma=1/\tau$ is the elastic relaxation rate of electrons and $\Gamma_{T}$
is the part due to fluctuating defects. The ratio $\Gamma_{T}/\Gamma\approx T\kappa$
is roughly linear in temperature and is determined by $\kappa$, the
relative density of states of thermally fluctuating defects that is
only weakly material dependent\cite{Yu1988}. The remaining parameters
that determine the strength of the fluctuations $p_{F}\ell$ and $\zeta$
can be determined from independent measurements. The noise amplitude
grows rapidly as the device becomes smaller and more disordered. As
a result, this mechanism is likely to be the dominant source of inductance
noise in small and highly disordered devices. Eq.~(\ref{3Dnoise})
holds for three dimensional wires with thickness $L_{\perp}>\zeta$,
we will see that the fluctuations get rather larger for very thin,
two and one dimensional, wires $\ell<L_{\perp}<\zeta$. 

Note that throughout this paper we distinguish the sample to sample
fluctuations of kinetic inductance defined as,

\begin{equation}
\frac{\left\langle \delta K^{2}\right\rangle }{\left\langle K\right\rangle ^{2}}\equiv\frac{\left\langle (K-\left\langle K\right\rangle )^{2}\right\rangle }{\left\langle K\right\rangle ^{2}}\approx C_{3}\frac{1}{(p_{F}\ell)^{4}}\frac{\zeta\ell^{2}}{V},\label{eq:StSInductance}
\end{equation}
where the angular brackets mean averaging over disorder realizations.
This quantity is the saturation value of the noise amplitude $\delta_{K}^{2}$
defined above when the two disorder configurations $u$ and $u'$
in Eq.~(\ref{3Dnoise}) are completely uncorrelated. In other words
all of the impurities have changed their positions in Eq.~(\ref{eq:StSInductance})
as opposed to only a fraction in Eq.~(\ref{3Dnoise}). In this way
Eq.~(\ref{eq:StSInductance}) is the upper limit of the noise amplitude
Eq.~(\ref{3Dnoise}). The result in the right hand part is valid
in three dimensional wires $\zeta\ll L_{x},\, L_{y},\, L_{z}$, and
we will see that $C_{3}\approx25$.

Up to a numerical factor, the results Eqs.~(\ref{3Dnoise})~and~(\ref{eq:StSInductance})
can be derived from the following qualitative argument. We first estimate
the amplitude of the sample-to-sample fluctuations. Consider a small
cubic piece of a superconductor of the size $L^{3}$ with $L\lesssim\zeta.$
At these scales the coherence of electrons is weakly affected by the
superconductivity. Optical sum rule $\int\sigma(\omega)d\omega=const$
and Anderson theorem of gap disorder independence imply that superfluid
response is directly related to normal state conductivity~\cite{DeGennes}.
Therefore the fluctuations of the normal state conductivity $\sigma$
roughly translate into fluctuations of the superfluid response, $\delta_{K}^{2}\sim\delta_{\sigma}^{2}$.
Conductance fluctuations of a small piece of metal have a universal
value $\left\langle \delta\sigma^{2}\right\rangle \approx\left(2e^{2}/h\right)^{2}$.
Thus, one expects that the maximal change in the interference pattern
correspond to the relative change of the superfluid response by $\left\langle \delta K^{2}\right\rangle /\left\langle K\right\rangle ^{2}=1/(p_{F}^{2}\ell L)^{2}$
in three dimensional wires.  This is the amplitude of sample-to-sample
fluctuations of a mesoscopic size piece of superconductor. We now
estimate the fluctuation of the superfluid response during the time
$t\gg1/\Delta,\, L^{2}/D$. During the time $t$ a small number of
defects change their position in space, which means that in contrast
to the sample-to-sample fluctuations only a small number of electronic
paths are affected, resulting in a smaller value of the fluctuation
of the superfluid response. We expect the fluctuation to be proportional
to the number of paths affected by the motion of the defects. A typical
path of a diffusive electron that enters and exits a cube of size
$L$ has a length $(L/\ell)^{2}\ell$, the probability that the concentration
$n_{T}$ of randomly positioned thermally activated impurities affect
this path is $\sigma_{T}n_{T}(L/\ell)^{2}\ell$ where $\sigma_{T}$
is the scattering cross-section of the fluctuating impurities. Assuming
that the fluctuating and static impurities are roughly equivalent
we can relate $n_{T}\sigma_{T}=\gamma_{T}n_{i}\sigma=\gamma_{T}/\ell$,
where $\sigma$ is the average scattering cross-section for all impurities
and $n_{i}$ is the total impurity concentration. Combining all these
factors together we get $\delta_{K}^{2}=\gamma_{T}/(p_{F}l)^{4}$.

We assume for the sake of this estimate that in a larger sample $L>\zeta$
the information of the single electron phases is lost on the scale
of the coherence length $\zeta$. This implies that the regions of
the size $\zeta$ fluctuate independently. Electromagnetic response
of the whole sample is obtained by adding these regions as independent
resistor network, adding $\zeta/L$ independent factors we get Eq.~(\ref{3Dnoise})
for the fluctuating part of the kinetic inductance. The fraction of
the paths affected has to be modified as well $\gamma_{T}(\zeta/\ell)^{2}$.
Combining these factors we obtain Eq.~(\ref{3Dnoise}). We will see
that the assumption of uncorrelated fluctuations on the scale of $\zeta$
is violated in the case of thin films and wires.

In the next Section~\ref{sec:Sample-to-sample} we calculate sample-to-sample
fluctuations of the superfluid response. After that we use this result
to calculate the noise amplitude in Section~\ref{sec:Noise-amplitude}
and conclude in Section~\ref{sec:Conclusion}. Details of construction
of diagramatic perturbation theory are given in Appendix~\ref{sec:Details-of-diagrams}.
Calculation of Fourier integrals of the superconducting coherence
factors is shown in Appendix~\ref{sec:FourierT0}.

\section{Sample-to-sample fluctuation of the superfluid response\label{sec:Sample-to-sample}}

We now turn to the analytical computation. We focus on the properties
of a superconducting wire with the simplest geometry: a rectangle
of total volume $V=L_{x}\times L_{y}\times L_{z}$ connected to two
leads that carry spatially uniform supercurrent. A small supercurrent
is injected into the wire, along $Oz$ axis, by an external source
and the resulting phase difference is measured, for example with Josephson
junctions. This geometry is very similar to the one used in Ref.~\onlinecite{McDermott}
to measure SQUID inductance. Local electro-magnetic response of a
superconductor is given by the kernel defined as $j_{\alpha}(r)=\int dr'S_{\alpha\beta}(r,r')A_{\beta}(r')$.
This supercurrent response can be thought of as a superfluid density,
$S$. Since the definition of the kernel is a local form of London
equation. Kinetic inductance, $K$, is a result of the work done by
electromagnetic field to accelerate the Cooper pairs and is therefore
inversely proportional to the superfluid density response $S$, $K\sim1/S$.
Therefore $\left\langle \delta K^{2}\right\rangle /\left\langle K\right\rangle {}^{2}\approx\left\langle \delta S^{2}\right\rangle /\left\langle S\right\rangle {}^{2}$
and in the following we will discuss fluctuations of $S$. The fluctuation
in the total response of the wire of the volume $V$ is given by the
spatial average, 
\begin{equation}
\langle\delta S^{2}\rangle=\int\frac{\prod_{i=1}^{4}d\mathbf{r_{i}}}{L_{z}^{4}}\left[\left\langle S(\mathbf{r_{1},r_{2}})S(\mathbf{r_{3},r_{4}})\right\rangle -\left\langle S\right\rangle ^{2}\right].
\end{equation}

We introduce the exact single particle eigenstates of the disordered
system, 
\begin{eqnarray*}
\left(-\frac{1}{2m}\nabla^{2}-\mu+u(\mathbf{r})\right)\phi_{\xi}(r) & = & \xi\phi_{\xi}(r),
\end{eqnarray*}
where we assumed Gaussian delta-correlated disorder $\langle u(\mathbf{r})u(\mathbf{r'})\rangle=n_{i}u_{0}^{2}\delta(\mathbf{r-r'})$
in the wire characterized by a momentum relaxation rate $\tau^{-1}=2\pi\nu n_{i}u_{0}^{2}$,
where $\nu$ is the density of single electron states and $n_{i}$
density of impurities. We assume that $\langle u(\mathbf{r})\rangle=0$
without loss of generality as non-zero value would result in a shift
of the chemical potential $\mu$ that can be absorbed into its definition.
The response kernel of a superconductor to electromagnetic field can
be expressed in terms of these eigenstates, $\phi_{\xi}(r)$\cite{DeGennes},
\begin{gather}
S_{\alpha\beta}(r,r')=-\frac{e^{2}}{4m^{2}}\int d\xi_{1}d\xi_{2}p_{\xi_{1}\xi_{2}}^{\alpha}(r)p_{\xi_{1}\xi_{2}}^{\beta}(r')\mathcal{L}_{\xi_{1}\xi_{2}},\label{SinEigenStates}\\
\mathbf{p}_{\xi_{1}\xi_{2}}(r)\equiv\phi_{\xi}(r)\overleftrightarrow{\nabla}\phi_{\xi'}(r)=\phi_{\xi}(r)\nabla\phi_{\xi'}(r)-\phi_{\xi'}(r)\nabla\phi_{\xi}(r).\label{eq:Momentum}
\end{gather}
where we introduced a coherence factor\cite{BCS}, 
\begin{eqnarray}
 & \mathcal{L}_{\xi\xi'}=-\left[\frac{1}{2}\frac{EE'-\Delta^{2}-\xi\xi'}{EE'(E+E')}\left(1-f_{E}-f_{E'}\right)\right.\nonumber \\
 & \left.-\frac{1}{2}\frac{f_{E}-f_{E'}}{E-E'}\frac{EE'+\Delta^{2}+\xi\xi'}{EE'}-\frac{f_{\xi'}-f_{\xi}}{\xi-\xi'}\right].\label{mathcalL}
\end{eqnarray}
Here $E\equiv\sqrt{\xi^{2}+\Delta^{2}}$ and $f_{\xi}$ is the Fermi-Dirac
distribution. The 3rd term in Eq.~(\ref{mathcalL}) is obtained taking
$\Delta\rightarrow0$ in the first two terms. This term represents
the normal state diamagnetic part of the response function such that
the right hand side of Eq.~(\ref{SinEigenStates}) vanishes in the
normal state. Note that taking $\Delta\rightarrow0$, leads to $E\rightarrow|\xi|$
whereas the eigenstate energy $\xi\rightarrow\pm|\xi|$ can be both
positive and negative, and one has to keep in mind that $f_{-|\xi|}=1-f_{|\xi|}$.

It is useful to rewrite Eq.~(\ref{SinEigenStates}) in terms of exact
single particle Green functions, 
\begin{equation}
G_{\xi}^{R/A}(\mathbf{r,r'})=\sum_{n}\frac{\phi_{n}(\mathbf{r})\phi_{n}^{\ast}(\mathbf{r'})}{\xi-\xi_{n}\pm i\delta},
\end{equation}
as follows, 
\begin{equation}
\overline{S}=\frac{e^{2}}{4m^{2}}\int\frac{d\mathbf{r}d\mathbf{r'}d\xi d\xi'}{(2\pi)^{2}L_{z}^{2}}\mathcal{L}_{\xi\xi'}\Delta G_{\xi}\overleftrightarrow{\nabla}\overleftrightarrow{\nabla'}\Delta\overline{G}_{\xi'},\label{eq:ScDens}
\end{equation}
where we introduced a notation, $\Delta G_{\xi}\equiv G_{\xi}^{R}(\mathbf{r,r'})-G_{\xi}^{A}(\mathbf{r,r'})$
and $\Delta\overline{G}_{\xi}\equiv G_{\xi}^{R}(\mathbf{r',r})-G_{\xi}^{A}(\mathbf{r',r})$,
and the meaning of the double arrowed gradient symbol can be inferred
from Eq.~(\ref{eq:Momentum}).

The variance of the superconducting density, 
\begin{equation}
\langle\delta S^{2}\rangle=\int d\xi_{1}d\xi_{2}d\xi_{3}d\xi_{4}\mathcal{L}_{\xi_{1}\xi_{2}}\mathcal{L}_{\xi_{3}\xi_{4}}\mathcal{F}\label{eq:deltaS}
\end{equation}
using Eq.~(\ref{SinEigenStates}) can be written in terms of a correlator,
\begin{eqnarray}
 & \mathcal{F}\equiv\frac{W^{2}}{16}\left[\langle p_{\xi_{1}\xi_{2}}^{\alpha}p_{\xi_{1}\xi_{2}}^{\beta}p_{\xi_{3}\xi_{4}}^{\alpha'}p_{\xi_{3}\xi_{4}}^{\beta'}\rangle-\langle p_{\xi_{1}\xi_{2}}^{\alpha}p_{\xi_{1}\xi_{2}}^{\beta}\rangle\langle p_{\xi_{3}\xi_{4}}^{\alpha'}p_{\xi_{3}\xi_{4}}^{\beta'}\rangle\right],\label{eq:CorrelatorF}\\
 & W\equiv\frac{e^{2}}{m^{2}}\frac{1}{(2\pi)^{2}L_{z}^{2}}.\nonumber 
\end{eqnarray}

\begin{figure}[t]
\includegraphics[width=0.9\columnwidth]{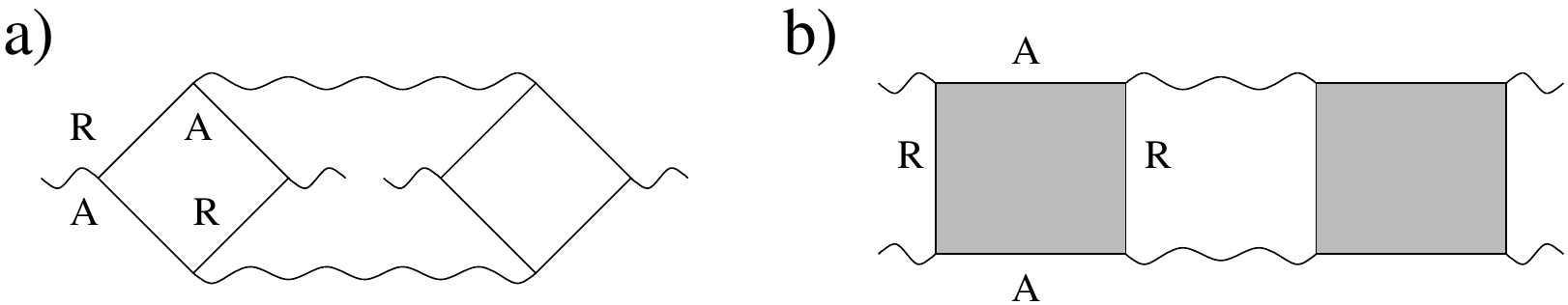}\caption{(a), (b) Diagrams contributing to mesoscopic sample-to-sample fluctuations
of superconducting density. Solid lines represent disorder averaged
Green function; short wavy lines stand for current vertexes. The shaded
regions in (b) indicate appropriately renormalized Hikami boxes, see
text and Fig.~\ref{fig:Hikami}. Long wavy lines stand for diffuson/Cooperon
impurity ladders $\mathcal{D}(x_{1},x_{2})$, see discussion after
Eq.~(\ref{eq:DiamondDiagramTwoT}).}

\label{fig:AARRDiagrams} 
\end{figure}

A similar correlator to $\mathcal{F}$ appears in the variance of
conductance fluctuations in the normal metal for which a perturbation
theory in $k_{F}\ell\gg1$ was developed and used extensively\cite{LeeStoneFukuyama}.
The main order contribution in this perturbation series may be written
in terms of the diagrams shown in Fig.~\ref{fig:AARRDiagrams}. In
the case of a superconductor the diagrams will have a similar form. 

\begin{figure}
\includegraphics[width=0.85\columnwidth]{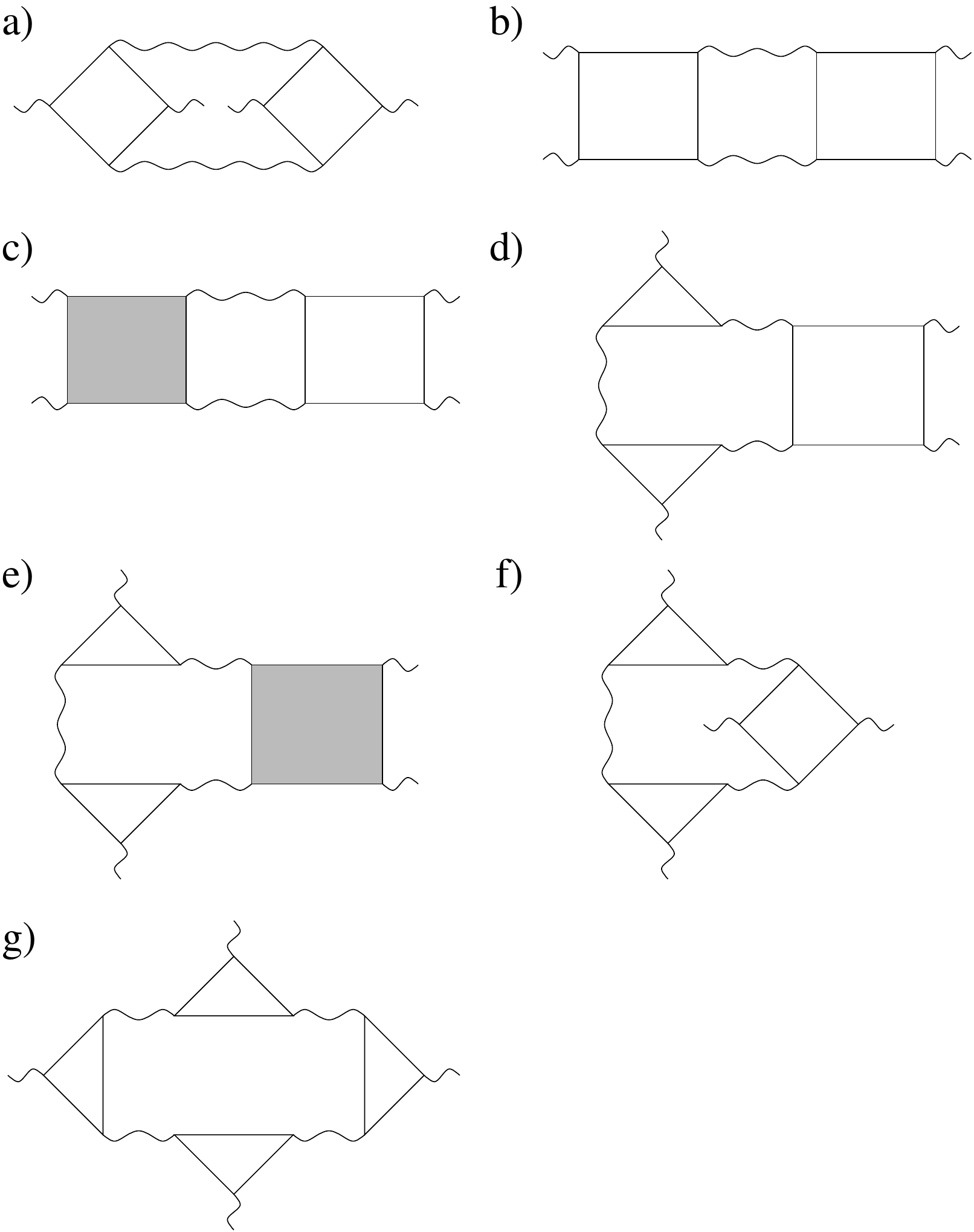}

\caption{Diagrams obtained originating from the contributions of the form $G^{R}(x_{1},x_{2})G^{R}(x_{2},x_{1})$
to each of the superfluid densities Eq.~(\ref{eq:ScDens}) in the
correlator Eq.~(\ref{eq:deltaS}). These diagrams cancel in the case
of normal state conductance calculation yet contribute to the superfluid
density.}

\label{fig:34diags}
\end{figure}

It is instructive therefore to compare expressions Eqs.~(\ref{eq:ScDens},~\ref{eq:deltaS},~\ref{eq:CorrelatorF})
with the normal case. The amplitude of normal state conductance fluctuations
contains only the diagrams shown in Fig.~\ref{fig:AARRDiagrams}.
It has been shown rigorously\cite{BarangerStone} that dissipative
normal state conductance is fully determined by products of one retarded
and one advanced Green functions averaged together $G^{R}G^{A}$,
compare Eq.~(\ref{eq:ScDens}). Therefore expanding the normal state
analog of the Eq.~(\ref{eq:ScDens}) in terms of $G^{R}$ and $G^{A}$
(in other words the formula for conductance) the products of two retarded
or two advanced Green functions in Eq.~(\ref{eq:ScDens}) that describe
non-dissipative diamagnetic currents vanish. A normal state analog
of the correlator $\mathcal{F}$ in Eq.~(\ref{eq:CorrelatorF}) contains
therefore only the averages of the type $\langle G^{R}G^{A}G^{R}G^{A}\rangle$
resulting in the diagrams in Fig.~\ref{fig:AARRDiagrams} being the
only diagrams contributing to conductance fluctuations. In contrast
to the normal phase, the superfluid density is a thermodynamic property
of a superconductor and despite Anderson theorem relating it to the
normal state conductivity the same arguments do not apply. One has
to be cautious and take into account a number of additional diagrams
shown in Fig.~\ref{fig:34diags} which vanish in a normal state yet
give non-zero contribution in the superconducting phase. These diagrams
originate form the terms of the form $G^{R}G^{R}$ in the superconducting
density Eq.~(\ref{eq:ScDens}). A more detailed discussion of the
standard procedure of constructing the diagramatic perturbation theory
can be found in Appendix~\ref{sec:Details-of-diagrams}.

The diagrams in Figs.~\ref{fig:AARRDiagrams} and \ref{fig:34diags}
can be simplified in the standard way\cite{LeeStoneFukuyama,AkkermansBook}
by identifying two types of blocks characterized by distinct length
scales: the mean free path $\ell$ and the coherent diffusion length
scale $\zeta\gg\ell$. For the diagram in Fig.~\ref{fig:AARRDiagrams}(a)
such separation is done as follows, 
\begin{eqnarray}
 & \mathcal{F}_{1a}\equiv W^{2}\int\prod_{i=1}^{4}dx_{i}j(x_{1},x_{2})j(x_{3},x_{4})\mathcal{D}(x_{1},x_{3})\mathcal{D}(x_{4},x_{2})\nonumber \\
 & \times\exp(i\xi_{1}t_{1}-i\xi_{3}t_{3}+i\xi_{4}t_{4}-i\xi_{2}t_{2}),\label{eq:DiamondDiagram}
\end{eqnarray}
where $x_{i}\equiv(\mathbf{r_{i}},t_{i})$. We introduced the 'Hikami
boxes' $j(x_{1},x_{2})$ and the diffuson/Cooperon propagators $\mathcal{D}(x_{1},x_{2})$,
see Fig.~\ref{fig:AARRDiagrams}. In the main order in $p_{F}\ell\gg1$
and $\zeta/\ell\gg1$ these blocks can be averaged over disorder independently
of each other.

\begin{figure}[t]
\includegraphics[width=0.5\columnwidth]{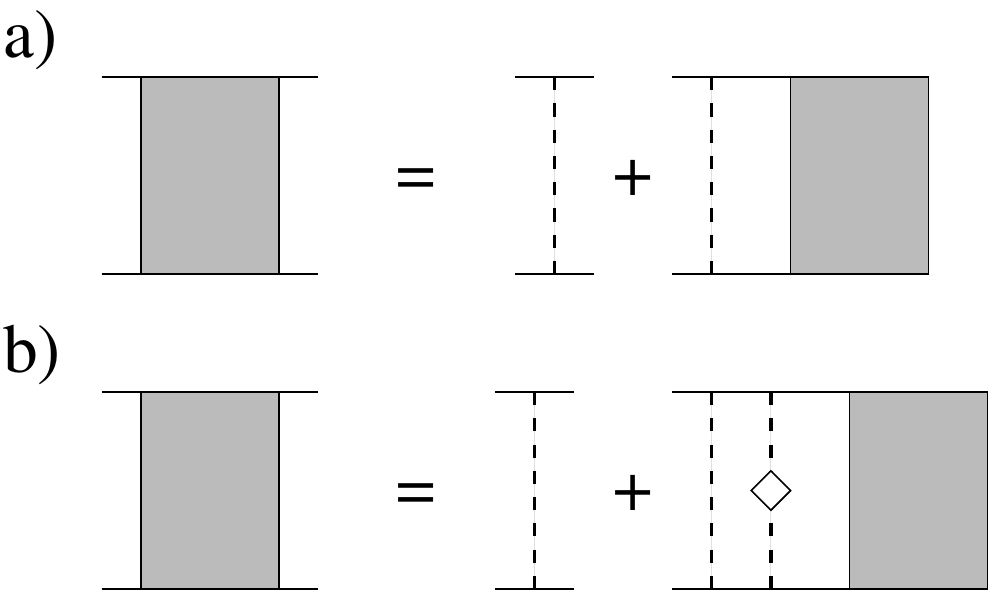}

\caption{Dyson equations for diffusons/Cooperons (shaded regions) in time reversal
symmetric case. Vertical dashed lines stand for the disorder correlator
$\langle u(\mathbf{r})u(\mathbf{r'})\rangle=\frac{1}{2\pi\nu\tau}\delta(\mathbf{r-r'})$,
solid lines correspond to disorder averaged Green functions. (a) Corresponds
to static impurity configuration. (b) Corresponds to the case when
a fraction of impurities have moved between different measurements.
The additional ``diamond'' vertex corresponds to the correlator
$\langle\left(u_{\mathbf{r_{i}}+\delta\mathbf{r}}-u_{\mathbf{r_{i}}}\right)u_{\mathbf{r_{i}}}\rangle$.}

\label{fig:Dyson+TLS} 
\end{figure}

The diffusion propagator is defined as a joint average of two Green
functions $\mathcal{D}(x_{1},x_{2})=\frac{1}{2\pi\nu\tau^{2}}\langle G^{R}(x_{1},x_{2})G^{A}(x_{2},x_{1})\rangle$
and describes the long-range coherent diffusion of electrons on the
scales $|\mathbf{r_{1}}-\mathbf{r_{2}}|\gg\ell$\cite{AkkermansBook}.
The diffuson propagators are represented diagramatically by impurity
ladders of diffuson and Cooperon type\cite{AkkermansBook} and satisfy
Dyson-like equations Figs.~\ref{fig:Dyson+TLS}(a). In the diffusive
regime Figs.~\ref{fig:Dyson+TLS}(a) reduces to the standard diffusion
equation. To model a realistic conductor this diffusion equation has
to be supplemented with boundary conditions. We set $\mathcal{D}=0$
at the contacts and $\nabla\mathcal{D}=0$ at the surface of the wire.
The diffusion propagators can be written in terms of the eigenmodes
of the diffusion equation $\Phi_{\mathbf{q}}(\mathbf{r})=\sqrt{\frac{2^{3}}{L_{x}L_{y}L_{z}}}\sin q_{x}x\cos q_{y}y\cos q_{z}z$,
$\mathbf{q}=\left[\frac{\pi n_{x}}{L_{x}},\frac{\pi n_{y}}{L_{y}},\frac{\pi n_{z}}{L_{z}}\right]$,
$n_{x}=1,2,..$ $n_{y},n_{z}=0,1,2...$, 
\begin{gather}
\mathcal{D}(t,\mathbf{r,r'})=\frac{1}{2\pi\nu\tau^{2}}\sum_{q}e^{-Dq^{2}t}\Phi_{q}(\mathbf{r})\Phi_{q}(\mathbf{r'}),
\end{gather}
where $t\geq0$ stands for the diffusion time, $D$ is the diffusion
coefficient for electrons and $\nu$ is the density of electronic
states. Averaging over the wire volume in Eq.~(\ref{eq:DiamondDiagram})
of the product of orthogonal eigenfunctions $\Phi_{\mathbf{q}}(\mathbf{r})$
gives rise to the momentum conservation condition (and this is true
also for all other diagrams). As a result all diffusion propagators
in each diagram in Figs.~\ref{fig:AARRDiagrams},~\ref{fig:34diags}
have the same momentum $q$. Sums over $q$ may be approximated by
integral in the case $\zeta/L_{i}\ll1,\, i=x,\, y,\, z$, 
\begin{equation}
\sum_{q}\rightarrow\frac{\Omega_{d}}{\pi^{d}\zeta^{d}}\int_{0}^{\infty}dk_{i},
\end{equation}
where $\Omega_{d}=\left[V,\, L_{z}L_{y},\, L_{z}\right]$ for $d=3,2,1$.
We define dimensionality $d$ of the sample with respect to the coherence
length, i.e. $d=3$ corresponds to $L_{x},\, L_{y},\, L_{z}\gg\zeta$,
$d=2$ corresponds to $L_{x},\, L_{z}\gg\zeta>L_{y}$ and $d=1$ corresponds
to $L_{z}\gg\zeta>L_{x},\, L_{y}$. 

The following relation will be useful, 
\begin{equation}
\sum_{q}q^{2m}e^{-Dq^{2}t}\approx\frac{\kappa_{m}}{2^{d}\pi^{d/2}}\frac{1}{(t\Delta){}^{d/2+m}}\frac{\Omega_{d}}{\zeta^{d}}\frac{1}{\zeta^{2m}},\label{eq:IntOverQDiffuson}
\end{equation}
where $m=0,1,2$ and $\kappa_{m}=1,\frac{1}{2},\frac{3}{4}$ respectively.
In the absence of magnetic fields the Cooperon and diffuson propagators
are identical and we will not distinguish them in the following, instead
including a factor of $2$ in front of all the diagrams.

\begin{figure}
\includegraphics[width=0.8\columnwidth]{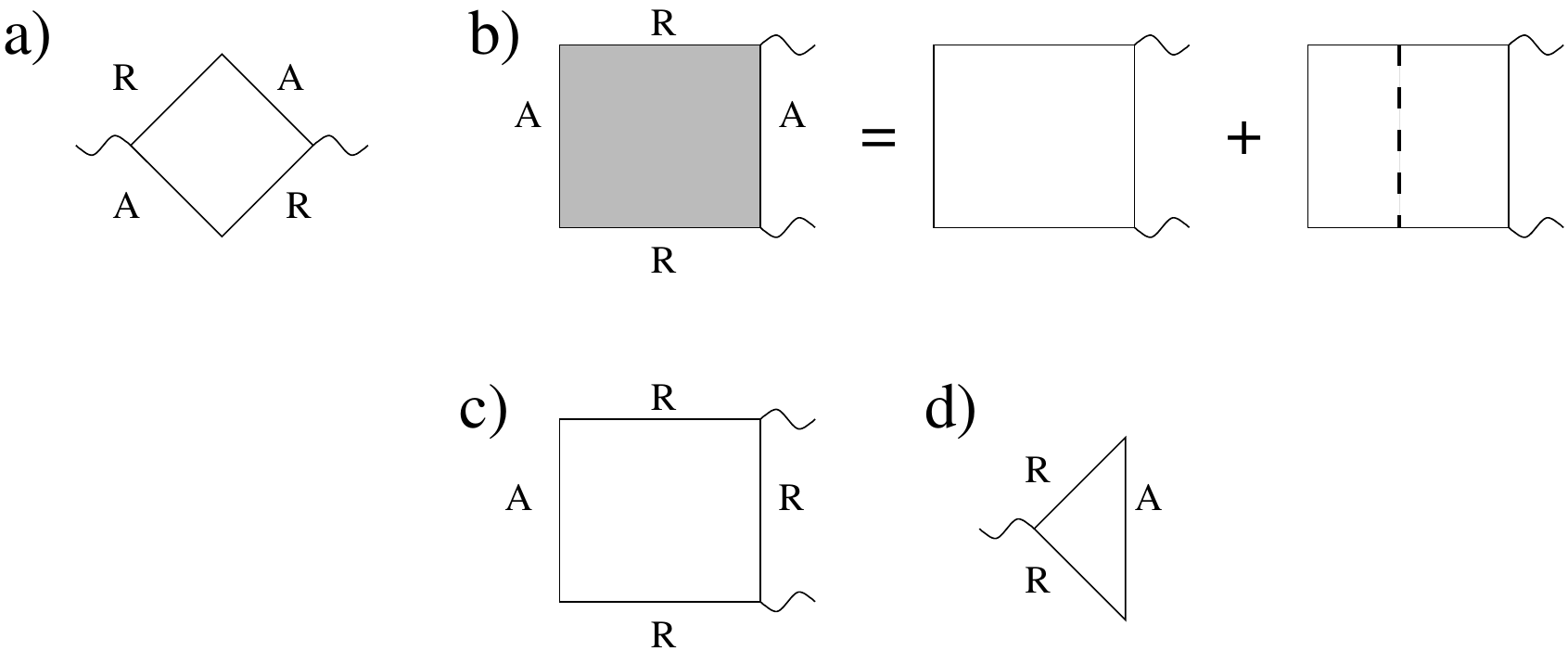}

\caption{Four types of Hikami boxes appearing in the diagrams in Figs.~\ref{fig:AARRDiagrams}
and \ref{fig:34diags}.}

\label{fig:Hikami}
\end{figure}

The second type of blocks that we introduced in Eq.~(\ref{eq:DiamondDiagram})
are Hikami boxes $j(x_{1},x_{2})$. There are four types shown in
Fig.~\ref{fig:Hikami}, where straight lines represent disorder averaged
Green functions. $j(x_{1},x_{2})$ decays exponentially on the short
length scales of the order of the mean free path $|r|\sim\ell\ll L_{x},\, L_{y}$
and therefore can be approximated by a delta function $j(x_{1},x_{2})\approx j_{0}\delta(x_{1}-x_{2})$.
The constant factor $j_{0}$ can be calculated in momentum space using
disorder averaged Green functions. The result of this calculation
for each of the vertexes in Fig.~\ref{fig:Hikami}(a)-(d) reads,
\begin{gather}
j_{a}\approx\frac{4\pi\nu\tau^{3}p_{F}^{2}}{d},\\
j_{b}\approx\frac{2\pi\nu\tau^{3}p_{F}^{2}}{d},\\
j_{c}\approx-\frac{2\pi\nu\tau^{3}p_{F}^{2}}{d},\\
j_{d}^{\alpha}\approx\pm\frac{4\pi\nu\tau^{3}p_{F}^{2}}{d}\frac{q_{\alpha}}{m},
\end{gather}
where the expression for the last vertex depends explicitly on the
momentum of the diffusion eigenmodes $\mathbf{q}$.

Using the short range character of the current vertexes we simplify
Eq.~(\ref{eq:DiamondDiagram}) for the diagram in Fig.~\ref{fig:AARRDiagrams}~(a),

\begin{eqnarray}
 & \mathcal{F}_{1a}\equiv W^{2}j_{a}^{2}\int d\mathbf{r}d\mathbf{r'}dtdt'\mathcal{D}(t,\mathbf{r},\mathbf{r'})\mathcal{D}(t',\mathbf{r'},\mathbf{r})\nonumber \\
 & \times\exp(-i(\xi_{1}-\xi_{4})t-i(\xi_{3}-\xi_{2})t').\label{eq:DiamondDiagramTwoT}
\end{eqnarray}
Note that all energy dependence in the above expression is contained
in Fourier factors. As a result after substituting Eq.~(\ref{eq:DiamondDiagramTwoT})
into Eq.~(\ref{eq:deltaS}) we can take integrals with respect to
energy in Eq.~(\ref{eq:deltaS}). The expression for the diagram
in Fig.~\ref{fig:AARRDiagrams}(a) therefore can be rewritten using
Eqs.~(\ref{eq:deltaS},\ref{eq:DiamondDiagramTwoT},\ref{eq:IntOverQDiffuson}),

\begin{gather}
\langle\delta S^{2}\rangle_{1a}=\frac{2^{-d}\pi^{-\frac{d}{2}}W^{2}j_{a}^{2}\Delta^{2}}{(2\pi\nu\tau^{2})^{2}}\frac{\Omega_{d}}{\zeta^{d}}\int_{0}^{\infty}\frac{d\lambda d\lambda'\left(\Theta_{\lambda+\lambda'}^{(-)}\right)^{2}}{(\lambda+\lambda')^{d/2}},\label{eq:DiamFullExp}
\end{gather}
and introducing the Fourier integral,

\begin{equation}
\Theta_{\lambda+\lambda'}^{(-)}\equiv\frac{1}{\Delta}\int d\xi d\xi'\mathcal{L}_{\xi\xi'}e^{i\xi\frac{\lambda}{\Delta}-i\xi'\frac{\lambda'}{\Delta}}=-2\pi\int_{\lambda+\lambda'}^{\infty}dxK_{0}(x),\label{eq:Fourier0}
\end{equation}
where $K_{n}(x),\, n=0,1,2,..$ is the modified Bessel function\cite{GR}
and we introduced a dimensionless time $\lambda\equiv t\Delta$ measured
in units of $1/\Delta$. See Appendix~\ref{sec:FourierT0} for details
of Fourier integration. Note that the integral over $\lambda,\,\lambda'$
in the right hand side of Eq.~(\ref{eq:DiamFullExp}) is dimensionless
and therefore simply represents a numerical coefficient.

It is instructive to show the calculation of the diagram in Fig.~\ref{fig:AARRDiagrams}(b),

\begin{eqnarray*}
 & j_{b}^{2}\int d\mathbf{r}d\mathbf{r'}\mathcal{D}(t,\mathbf{r},\mathbf{r'})\mathcal{D}(t',\mathbf{r'},\mathbf{r}))\\
 & \times\exp(-i(\xi_{1}-\xi_{4})t-i(\xi_{1}-\xi_{4})t').
\end{eqnarray*}
Including the coherence factors in Eq.~(\ref{eq:deltaS}) and using
Eq.~(\ref{eq:IntOverQDiffuson}) we get,

\begin{gather}
\langle\delta S^{2}\rangle_{1b}=\frac{2^{-d}\pi^{-\frac{d}{2}}W^{2}j_{b}^{2}\Delta^{2}}{(2\pi\nu\tau^{2})^{2}}\frac{\Omega_{d}}{\zeta^{d}}\int_{0}^{\infty}\frac{d\lambda d\lambda'}{(\lambda+\lambda')^{d/2}}\left(\Theta_{\lambda+\lambda'}^{(0)}\right)^{2},\label{eq:LDOSDiag}
\end{gather}
where crucially the Fourier integral has a distinct form from that
in Eq.~(\ref{eq:Fourier0}),

\begin{eqnarray}
 & \Theta_{\lambda+\lambda'}^{(0)}\equiv\frac{1}{\Delta}\int d\xi d\xi'\mathcal{L}_{\xi\xi'}e^{i\xi\frac{(\lambda+\lambda')}{\Delta}}\nonumber \\
 & =-\pi\int_{\lambda+\lambda'}^{\infty}dx\left(K_{0}(x)+\frac{1-xK_{1}(x)}{x^{2}}\right).\label{eq:ThetaZero}
\end{eqnarray}
Note that here we switched to dimensionless time $\lambda\equiv\Delta t$.
The details of the integration are shown in Appendix~\ref{sec:FourierT0}.
The rest of the diagrams can be computed in the analogous way noting
that the diagrams in Fig.~\ref{fig:34diags} contain the third type
of the Fourier transform of the coherence factor,

\begin{eqnarray}
 & \Theta_{\lambda+\lambda'}^{(+)}\equiv\frac{1}{\Delta}\int d\xi d\xi'\mathcal{L}_{\xi\xi'}e^{i\xi\frac{\lambda}{\Delta}+i\xi'\frac{\lambda'}{\Delta}}\nonumber \\
 & =-2\pi\int_{\lambda+\lambda'}^{\infty}dx\frac{1-xK_{1}(x)}{x^{2}}.\label{eq:ThetaPlus}
\end{eqnarray}
 Note that the long range asymptotic, $x\gg1$, of the Bessel function
is proportional to an exponent $K_{\nu}(x)\approx\sqrt{\frac{\pi}{2x}}e^{-x}$.
This means that the Fourier transform of the coherence factor Eq.~(\ref{mathcalL})
$\Theta_{\lambda}^{(-)}\sim e^{-\lambda}$ determines the roughly
exponential phase relaxation in the diagram in Fig.~\ref{fig:AARRDiagrams}(a),
Eq.~(\ref{eq:DiamFullExp}). This diagram corresponds to the fluctuations
in the transmission coefficient of electrons in the superconductor.
In contrast, in the diagram in Fig.~\ref{fig:AARRDiagrams}(b), Eq.~(\ref{eq:LDOSDiag}),
$\Theta_{\lambda}^{(0)}\sim\frac{1}{\lambda}$ and therefore the coherent
diffusion is only suppressed by a slow power law phase relaxation
rather than an exponent. An analogous slow relaxation of diffusion
modes is described by the diagrams in Fig.~\ref{fig:34diags}. Note
that the non-exponential relaxation of the diffusion modes is precisely
the reason why the diagrams in Fig.~\ref{fig:34diags} give non-zero
contribution in contrast to the normal metal case. This can be seen
by replacing each of the Fourier integrals $\Theta_{\lambda}^{(-)},\,\Theta_{\lambda}^{(0)},\,\Theta^{(+)}$
with an exponent $e^{-\lambda}$ in which case the sum of the diagrams
in Fig.~\ref{fig:34diags} vanishes, see Appendix~\ref{sec:Details-of-diagrams}
for details. The consequences of non-exponential relaxation is two-fold:
(i) due to both the contribution of the numerous additional diagrams
and slow-decaying non-exponential integrands the amplitude of the
superfluid density fluctuations is enhanced; (ii) as we will see in
the following the noise amplitude acquires non-linear dependence on
the effective impurity strength in one and two dimensions. 

Summing all the diagrams in Figs.~\ref{fig:AARRDiagrams},\ref{fig:34diags}
including the appropriate symmetry factors, see Appendix~\ref{sec:Details-of-diagrams},
we arrive at, 
\begin{gather}
\frac{\langle\delta S^{2}\rangle}{S^{2}}=C_{d}\frac{1}{(p_{F}\ell)^{4}}\frac{\zeta^{4-d}\ell^{2}\Omega_{d}}{V^{2}},\label{eq:FluctAmpResult}\\
C_{d}\equiv2^{1-d}9\pi^{-d/2}\left(\mathcal{I}^{RA}+2\mathcal{I}^{RR}\right).\label{eq:CoefficientUCF}
\end{gather}
where $\Omega_{d}=\left[V,L_{z}L_{y},L_{z}\right]$ for $d=3,2,1$.
The explicit form of the integrals in Eq.~(\ref{eq:CoefficientUCF})
reads, 
\begin{gather}
\mathcal{I}^{RA}=\int_{0}^{\infty}d\lambda_{1}d\lambda_{2}\frac{\left(\Theta_{\lambda_{1}+\lambda_{2}}^{-}\right)^{2}+\frac{1}{2}\left(\Theta_{\lambda_{1}+\lambda_{2}}^{0}\right)^{2}}{(\lambda_{1}+\lambda_{2})^{d/2}}A_{\lambda_{1}+\lambda_{2}},\label{eq:IRA}
\end{gather}
corresponding to the contribution of the diagrams in Fig.~\ref{fig:AARRDiagrams},
and

\begin{gather}
\mathcal{I}^{RR}=\int_{0}^{\infty}d\lambda_{1}d\lambda_{2}\frac{\left(\left(\Theta_{\lambda_{1}+\lambda_{2}}^{+}\right)^{2}+\frac{3}{2}\Theta_{\lambda_{1}+\lambda_{2}}^{+}\Theta_{\lambda_{1}+\lambda_{2}}^{0}\right)A_{\lambda_{1}+\lambda_{2}}}{(\lambda_{1}+\lambda_{2})^{d/2}}\nonumber \\
-2\int_{0}^{\infty}d\lambda_{1}d\lambda_{2}d\lambda_{3}\frac{\left(\Theta_{\lambda_{1}+\lambda_{2}+\lambda_{3}}^{+}\right)^{2}A_{\lambda_{1}+\lambda_{2}+\lambda_{3}}}{(\lambda_{1}+\lambda_{2}+\lambda_{3})^{d/2+1}}\nonumber \\
-2\int_{0}^{\infty}d\lambda_{1}d\lambda_{2}d\lambda_{3}\frac{\Theta_{\lambda_{1}+\lambda_{2}+\lambda_{3}}^{+}\Theta_{\lambda_{1}+\lambda_{2}+\lambda_{3}}^{0}A_{\lambda_{1}+\lambda_{2}+\lambda_{3}}}{(\lambda_{1}+\lambda_{2}+\lambda_{3})^{d/2+1}}\nonumber \\
+\int_{0}^{\infty}d\lambda_{1}d\lambda_{2}d\lambda_{3}d\lambda_{4}\frac{9\left(\Theta_{\lambda_{1}+\lambda_{2}+\lambda_{3}+\lambda_{4}}^{+}\right)^{2}A_{\lambda_{1}+\lambda_{2}+\lambda_{3}+\lambda_{4}}}{(\lambda_{1}+\lambda_{2}+\lambda_{3}+\lambda_{4})^{d/2+2}},\label{eq:IRR}
\end{gather}
corresponding to the contribution of the diagrams in Fig.~\ref{fig:34diags}.
The factor $A_{\lambda}=1$ in Eq.~(\ref{eq:FluctAmpResult}). We
estimate the coefficients in Eq.~(\ref{eq:FluctAmpResult}) by numerically
taking the integrals in Eqs.~(\ref{eq:IRA},~\ref{eq:IRR}) $C_{3}\approx25,C_{2}\approx60,C_{1}\approx200$.

\section{\label{sec:Noise-amplitude}Noise amplitude}

A model of tunneling two level (bistable) defect, i.e. an impurity
fluctuating between two spatial positions, is used to describe noise
in a wide range of materials. For simplicity we use this model to
give a precise meaning to the amplitude of the noise calculated here.
All of the results can be easily generalized for a more generic dynamics
of impurities~\cite{AltshulerSpivak}. In the presence of a single
bistable defect superfluid response demonstrates significant fluctuations.
The noise amplitude for a single defect is given by the correlator,
\begin{gather}
\langle\delta S_{\delta\mathbf{r}}^{2}\rangle=\langle\delta S^{2}\rangle-\langle\delta S_{\mathbf{r_{i}+\delta}\mathbf{r}}\delta S_{\mathbf{r_{i}}}\rangle.\label{S_correlator}
\end{gather}
The first term on the right hand side of Eq.~(\ref{S_correlator})
is simply the fluctuation amplitude Eq.~(\ref{eq:FluctAmpResult}),
$\delta S\equiv S-\langle S\rangle$. The second term corresponds
to the diagrams in Figs.~\ref{fig:AARRDiagrams}~and~\ref{fig:34diags}
with one of the response kernels $\delta S_{\mathbf{r_{i}+\delta}\mathbf{r}}$
containing the disorder configuration with one defect shifted from
$\mathbf{r_{i}}$ by a distance $\mathbf{\delta r}$, and the other
$\delta S_{\mathbf{r_{i}}}$ containing the bistable defect in its
original position $\mathbf{r_{i}}$. This shift introduces an effective
dephasing rate cutting off the diffusion modes. To show this we recalculate
the diffuson/Cooperon in the presence of the shifted bistable defect.
The Dyson equation for the diffuson/Cooperon has to be modified by
including an additional vertex shown in Fig.~\ref{fig:Dyson+TLS}(b)
as an impurity line with a diamond\cite{Feng86,Falko90}. This new
vertex corresponds to the correlator $\langle\left(u_{\mathbf{r_{i}}+\delta\mathbf{r}}-u_{\mathbf{r_{i}}}\right)u_{\mathbf{r_{i}}}\rangle$.
It is obtained in the perturbation expansion in the small parameter,
$\langle\left(u_{\mathbf{r_{i}}+\delta\mathbf{r}}-u_{\mathbf{r_{i}}}\right)u_{\mathbf{r_{i}}}\rangle/\langle u_{\mathbf{r_{i}}}u_{\mathbf{r_{i}}}\rangle$.
In momentum space the last term in Fig.~\ref{fig:Dyson+TLS}(c) reads,
\begin{equation}
u_{0}^{2}\int\frac{d\mathbf{p}d\mathbf{p'}}{(2\pi)^{6}}|G_{\mathbf{p}}^{R}|^{2}|G_{\mathbf{p'}}^{A}|^{2}\left(\cos\left[\mathbf{(p-p')\delta r}\right]-1\right)\mathcal{D}(\mathbf{q}),
\end{equation}
where we keep only the main contribution in the gradient expansion
of the diffusion mode. This gives, 
\begin{gather}
\left(-i\omega+Dq^{2}+\tau_{eff}^{-1}\right)\mathcal{D}(q)=1,\\
\tau_{eff}^{-1}=\frac{1}{\tau}\alpha(k_{F}\delta r),\\
\alpha(k_{F}\delta r)\equiv\left(1-\frac{\sin^{2}\left(k_{F}\delta r/2\right)}{(k_{F}\delta r/2)^{2}}\right).
\end{gather}
In the presence of more than one thermally activated defect their
contributions simply add up for small enough concentration of such
defects $\gamma_{T}$,

\begin{equation}
\tau_{eff}^{-1}\approx\gamma_{T}\frac{1}{\tau}\langle\alpha(k_{F}\delta r)\rangle,\label{eq:Dephasing}
\end{equation}
which includes an average over characteristics of bistable defects
$\langle\alpha(k_{F}\delta r)\rangle$. For defects characterized
by $k_{F}\delta r\gg1$, $\langle\alpha(k_{F}\delta r)\rangle\approx1$.
The effect of bistable defects on the vertex parts of the diagrams
$j(r_{1},r_{2})$ is small as $\ell/\zeta$, and therefore we neglect
it in the following. The main order amplitude of the noise in the
superfluid density Eq.~(\ref{S_correlator}) is given by the same
diagrams in Figs.~\ref{fig:AARRDiagrams},~\ref{fig:34diags}. However,
in the case of the second term in Eq.~(\ref{S_correlator}) we need
to include the dephasing effect of thermally activated defects Eq.~(\ref{eq:Dephasing}).
The result is,

\begin{gather}
\delta_{K}^{2}=\widetilde{C}_{d}\frac{1}{(p_{F}\ell)^{4}}\frac{\zeta^{4-d}\ell^{2}\Omega_{d}}{V^{2}},\\
\widetilde{C}_{d}\equiv9\pi^{-d/2}2^{1-d}\left(\mathcal{I}^{RA}(\tau_{eff})+2\mathcal{I}^{RR}(\tau_{eff})\right),\label{eq:IntegralNoise}
\end{gather}
where the integrals $\mathcal{I}^{RA}(\tau_{eff}),\,\mathcal{I}^{RR}(\tau_{eff})$
are given by Eq.~(\ref{eq:IRA},~\ref{eq:IRR}) with the factor
$A_{t}$ modified to include the effect of bistable defects, 
\begin{gather}
A_{\lambda}\equiv1-e^{-\lambda/(\Delta\tau_{eff})}.\label{exponent}
\end{gather}
The first and the second term in Eq.~(\ref{exponent}) correspond
to the first and the second terms in Eq.~(\ref{S_correlator}). The
factor $\Omega_{d}$ has the same meaning as in Eq.~(\ref{eq:FluctAmpResult}).

The amplitude of the noise in $d=3$ can be estimated by expanding
the exponent in Eq.~(\ref{exponent}), $A_{\lambda}\approx\frac{\lambda}{\Delta\tau_{eff}}$,
resulting in, 
\begin{gather}
\delta_{K}^{2}\approx60\gamma_{T}\frac{1}{(p_{F}\ell)^{4}}\frac{\zeta^{3}}{V},\label{eq:Noise3DR}
\end{gather}
where we took the integrals in Eq.~(\ref{eq:IntegralNoise}) numerically
to estimate the value of the coefficient.

In lower dimensions the simple expansion of the factor in Eq.~(\ref{exponent})
is not possible since the integral over $t$ diverges. Instead we
keep only the main asymptotic contribution in the parameter $\Delta\tau_{eff}=3\zeta^{2}/(\gamma_{T}\ell^{2})\gg1$
which gives,

\begin{equation}
\delta_{K}^{2}=\frac{297}{4\pi}\gamma_{T}\log\left[\frac{3\zeta^{2}}{\gamma_{T}\ell^{2}}\right]\frac{1}{(p_{F}\ell)^{4}}\frac{\zeta^{4}}{VL_{y}},\label{eq:Noise2DR}
\end{equation}
in effective two dimensional superconductor, $L_{y}\lesssim\zeta\ll L_{x},L_{z}$,
and

\begin{equation}
\delta_{K}^{2}=99\left(3\gamma_{T}\right)^{\frac{1}{2}}\frac{1}{(p_{F}\ell)^{4}}\frac{\zeta^{4}\ell}{VS_{\perp}},\label{eq:Noise1DR}
\end{equation}
in effective one dimension, $L_{x},L_{y}\lesssim\zeta\ll L_{z}$.
The results Eqs.~(\ref{eq:Noise3DR}-\ref{eq:Noise1DR}) are valid
only in the limit of very small relative concentration of bistable
defects $\gamma_{T}\ll1$ such that the noise amplitude is much smaller
than the sample-to-sample fluctuation Eq.~(\ref{eq:FluctAmpResult})
which sets the upper limit on the noise amplitude. At high concentration
of fluctuating defects the noise amplitude saturates at the value
of sample-to-sample fluctuations Eq.~(\ref{eq:FluctAmpResult}).

\section{Conclusion\label{sec:Conclusion}}

We have performed the detailed analysis of kinetic inductance fluctuations
caused by motion of impurities in the superconductor and enhanced
by the mesoscopic quantum interference. We found that the effect is
closely related to the universal conductance fluctuations in normal
metals with an important distinction that the superconducting coherence
length determines the scale of coherent diffusion, in contrast to
the inelastic scattering length in the normal metal. We found that
the phase relaxation in superconductors has slow non-exponential character
which results in the enhanced amplitude of the noise and non-linear
scaling of the amplitude with the density of fluctuating impurities.
Our estimates of the magnitude of the noise suggest that the effect
is likely to dominate inductance noise in small devices. Experimentally,
the interference contribution to the inductance noise can be unambiguously
identified by driving supercurrents close to critical in magnitude
through the wire. In presence of strong supercurrent the superconducting
order parameter phase changes by $2\pi$ on the scale of the coherence
length which results in strong suppression of the quantum interference
and therefore the noise amplitude. 

The upper limit of the noise amplitude is given by the mesoscopic
fluctuation amplitude. We estimate this amplitude for an Al wire with
dimensions $100\textrm{nm}\times100\textrm{nm}\times10\mu\textrm{m}$
using Eq.~(\ref{eq:FluctAmpResult}), $\frac{\sqrt{\left\langle \delta K^{2}\right\rangle }}{K}\sim10^{-5}$. 

In normal state metals the mesoscopic noise amplitude was found to
be roughly temperature independent. This was attributed to a rough
cancellation of the temperature dependence of the density of thermally
activated defects $\gamma_{T}\propto T$ and the temperature smearing
of the mesoscopic fluctuation which roughly suppresses the amplitude
as $\propto1/T$~\cite{Birge1989}. In contrast, in the very low
temperature regime of transport in disordered superconductors considered
here the mesoscopic noise is limited by the superconducting coherence
length which demonstrates very weak temperature dependence away from
the transition temperature. As a result we expect the mesoscopic noise
in the kinetic inductance of disordered superconducting wires in presence
of thermally activated impurity dynamics to scale linearly with with
temperature $\propto T$ in the effective $d=3$ case and $\propto T\log T$
in $d=2$ and $\propto\sqrt{T}$ in $d=1$. This prediction could
be tested experimentally in superconducting thin films and wires.
\begin{acknowledgments}
The author is grateful to Lev Ioffe for motivating author's interest
in this problem and numerous comments and discussions. The author
is also grateful to Vladimir Fal'ko, Robert Smith, and Igor Lerner
for helpful discussions. This work is supported by LPS-CMTC. 
\end{acknowledgments}
%\bibliographystyle{apsrev4-1}
%\bibliography{Feng,NoiseInSuperconductingCircuits}


\appendix

\section{Details of the diagrams evaluation\label{sec:Details-of-diagrams}}

\begin{table}[h]
\begin{tabular}{|c|c|c|c|c|c|c|c|c|c|}
\hline 
Fig. & 1a & 1b & 2a & 2b & 2c & 2d & 2e & 2f & 2g\tabularnewline
\hline 
\hline 
factor & 2 & 4 & 4 & 8 & 8 & 8 & 16 & 8 & 12\tabularnewline
\hline 
\end{tabular}

\caption{Combinatorial factors corresponding to diagrams of different topologies
shown in Fig.~\ref{fig:AARRDiagrams},~\ref{fig:34diags} contributing
to the kinetic inductance fluctuations and the noise amplitude.}

\label{table}
\end{table}

Fluctuations of the superfluid response are given by the disorder,
spatial and energy average of a correlator $\mathcal{F}$ (see Eq.~(\ref{eq:deltaS}))
of four matrix elements of the current operator. This correlator can
be rewritten in terms of four imaginary parts of the exact Green functions,
see Eq.~(\ref{eq:ScDens}),

\begin{eqnarray}
 & \mathcal{F}\propto\left\langle \Delta G\overleftrightarrow{\nabla}\overleftrightarrow{\nabla'}\Delta\overline{G}\Delta G\overleftrightarrow{\nabla}\overleftrightarrow{\nabla'}\Delta\overline{G}\right\rangle \label{eq:SvImG}
\end{eqnarray}
The diagrams in Figs.~\ref{fig:AARRDiagrams},~\ref{fig:34diags}
are constructed by pairing Green functions to form diffuson and Cooperon
ladders, which are products of one retarded and one advanced Green
functions averaged over disorder $\left\langle G^{R}G^{A}\right\rangle $.
All possible such connections give rise to the diagrams in Figs.~\ref{fig:AARRDiagrams},~\ref{fig:34diags}.
In particular, the average of the form,

\begin{equation}
\left\langle G^{R}\overleftrightarrow{\nabla}\overleftrightarrow{\nabla'}G^{A}G^{R}\overleftrightarrow{\nabla}\overleftrightarrow{\nabla'}G^{A}\right\rangle \label{eq:GRGA}
\end{equation}
gives rise to the two diagrams shown in Fig.~\ref{fig:AARRDiagrams}(a),(b).
These are the diagrams that correspond to normal state conductance
fluctuations\cite{LeeStoneFukuyama,AkkermansBook,BarangerStone}.
The averages of the form,

\begin{equation}
\left\langle G^{A}\overleftrightarrow{\nabla}\overleftrightarrow{\nabla'}G^{A}G^{R}\overleftrightarrow{\nabla}\overleftrightarrow{\nabla'}G^{R}\right\rangle 
\end{equation}
give rise to the diagrams in Figs.~\ref{fig:34diags}(a),(b),(d),(f)
and (g). Finally, the averages of the form, 

\begin{equation}
-\left\langle G^{R}\overleftrightarrow{\nabla}\overleftrightarrow{\nabla'}G^{A}G^{R}\overleftrightarrow{\nabla}\overleftrightarrow{\nabla'}G^{R}\right\rangle 
\end{equation}
give rise to the diagrams shown in Fig.~\ref{fig:34diags}(c) and
(e). Crucially, the latter expression comes with a different sign
from expanding the expression in Eq.~(\ref{eq:SvImG}). 

Each diagram in Figs.~\ref{fig:AARRDiagrams} and \ref{fig:34diags}
comes with a different ``combinatorial factor'' reflecting the number
of times it occurs in the perturbation expansion of $\left\langle \delta S^{2}\right\rangle $.
This factor counts the number of ways to choose pairs of $\langle G^{R}G^{A}\rangle$
such that the two Green functions are parts of different superfluid
density loops. The corresponding factors are summarized in Table~\ref{table}.

In the case of mesoscopic fluctuations of the normal state conductance
it has been shown\cite{BarangerStone} that only the diagrams constructed
from $G^{R}G^{A}$ loops (Eq.~(\ref{eq:GRGA}) and Fig.~\ref{fig:AARRDiagrams})
give non-zero contribution. Since the symmetry factors and the Hikami
boxes in all of the diagrams constructed using $G^{R}G^{R}$ loops
(Fig.~\ref{fig:34diags}) are the same as in the normal state we
expect exact cancellation of the contribution of such diagrams to
the normal state conductance fluctuations. To very that this contribution
indeed vanishes we set all Fourier transformed coherence factors to
be equal to an exponent $\Theta_{\lambda}^{(0)},\Theta_{\lambda}^{(+)},\Theta_{\lambda}^{(-)}\rightarrow e^{-\lambda/\tau_{\varphi}}$
describing the effect of decoherence due to inelastic electron-electron
or electron-phonon collisions operational in the normal state. The
resulting sum of the integrals vanishes exactly $\mathcal{I}{}^{RR}(\tau_{\varphi})=0$.
Note that the non-exponential decay of $\Theta_{\lambda}^{(0)},\Theta_{\lambda}^{(+)},\Theta_{\lambda}^{(-)}$
is precisely the reason for non-vanishing contribution of the diagrams
in Fig.~\ref{fig:34diags} to the superfluid response fluctuations.

\section{Fourier integrals of the coherence factor \label{sec:FourierT0}}

\subsection{First type integral}

In the following we calculate the Fourier integral, 
\begin{gather}
\Theta_{t+t'}^{(-)}=\frac{1}{\Delta}\int_{-\infty}^{\infty}d\xi d\xi'\delta\mathcal{L}_{\xi\xi'}e^{-i\xi t+i\xi't'},
\end{gather}
in Eq.~(\ref{eq:DiamFullExp}), see also Ref.~\onlinecite{BCS}.
Note that in this appendix we perform the calculations using the real
time as opposed to the dimensionless time $\lambda$. We introduce
the latter at the end of the calculation. We introduce notations symmetric
with respect to $\xi\rightarrow-\xi$,

\begin{equation}
\mathcal{L}_{\xi\xi'}=\frac{1}{2}\frac{F(\xi)-F(\xi')+\xi\xi'\left[G(\xi)-G(\xi')\right]}{\xi^{2}-\xi'^{2}}.\label{eq:LvFG}
\end{equation}
where the symmetric functions $F(\xi)=F(-\xi)$ and $G(\xi)=G(-\xi)$
are defined as,

\begin{eqnarray}
 & F(\xi)=\left(E+\frac{\Delta^{2}}{E}\right)\left[1-2f(E)\right]-\xi\left[1-2f(\xi)\right],\label{eq:F}\\
 & G(\xi)=\frac{1}{E}\left[1-2f(E)\right]-\frac{1}{\xi}\left[1-2f(\xi)\right],\label{eq:G}
\end{eqnarray}
note that $f(-\xi)=1-f(\xi)$. 

Using Eqs.~(\ref{eq:LvFG},\ref{eq:F},\ref{eq:G}) we write, 
\begin{eqnarray}
 & \Theta_{t+t'}^{(-)}=\frac{2}{\Delta}\int_{0}^{\infty}d\xi d\xi'\frac{\cos\xi t\cos\xi't'\left[F(\xi)-F(\xi')\right]}{\xi^{2}-\xi'^{2}}\nonumber \\
 & +\frac{2}{\Delta}\int_{0}^{\infty}d\xi d\xi'\frac{\xi\xi'\sin\xi t\sin\xi't'\left[G(\xi)-G(\xi')\right]}{\xi^{2}-\xi'^{2}}\label{eq:ThetaMinusSym}
\end{eqnarray}
We replace the integral in the above expression by two principal value
(at point $\xi=\xi'$) integrals, 
\begin{gather}
A_{1}=2\lim_{a\rightarrow0,b\rightarrow\infty}\mathcal{P}'\int_{a}^{b}\int_{a}^{b}d\xi d\xi'\frac{\cos\xi t\cos\xi't'F(\xi)}{\xi^{2}-\xi'^{2}},\label{eq:A1even}\\
A_{2}=2\lim_{a\rightarrow0,b\rightarrow\infty}\mathcal{P}'\int_{a}^{b}\int_{a}^{b}d\xi d\xi'\frac{\xi\xi'\sin\xi t\sin\xi't'G(\xi)}{\xi^{2}-\xi'^{2}}.\label{eq:A1odd}
\end{gather}
Since $F(\xi)\rightarrow0$ as $\xi\rightarrow\infty$ the value of
the integral does not depend on the way the upper limit is approached.
Therefore we can set $b\rightarrow\infty$ right away.  Care must
be taken with the lower limit, for the integral in Eq.~(\ref{eq:A1even}),
\begin{gather}
\int_{a}^{\infty}d\xi'\frac{\cos\xi't'}{\xi^{2}-\xi'^{2}}=\int_{0}^{\infty}d\xi'\frac{\cos\xi't'}{\xi^{2}-\xi'^{2}}-\int_{0}^{a}d\xi'\frac{1}{\xi^{2}-\xi'^{2}},\label{eq:PVint}
\end{gather}
where in the last expression we assumed $a$ small enough so that
$\cos\xi't\approx1$. We evaluate the second term in Eq.~(\ref{eq:PVint}),
\begin{gather}
\int_{0}^{a}d\xi'\frac{1}{\xi^{2}-\xi'^{2}}=\frac{1}{2\xi}\ln\frac{\xi+a}{\xi-a},
\end{gather}
which is possible since $\xi>a$. In the first term in Eq.~(\ref{eq:PVint})
the principal value integral gives, 
\begin{gather}
\mathcal{P}'\int_{0}^{\infty}d\xi'\frac{\cos\xi't'}{\xi^{2}-\xi'^{2}}=\frac{\pi}{2}\frac{\sin\xi t'}{\xi},
\end{gather}
A similar integral in Eq.~(\ref{eq:A1odd}) gives, 
\begin{gather}
\mathcal{P}'\int_{0}^{\infty}d\xi'\frac{\xi'\sin\xi't'}{\xi^{2}-\xi'^{2}}=-\frac{\pi}{2}\cos\xi t'.
\end{gather}
So that 
\begin{eqnarray}
 & A_{1}+A_{2}=-\lim_{a\rightarrow0}\int_{a}^{\infty}\frac{d\xi}{\xi}F(\xi)\ln\frac{\xi+a}{\xi-a}\\
 & +2\lim_{a\rightarrow0}\int_{a}^{\infty}d\xi\cos\xi tF(\xi)\frac{\pi}{2}\frac{\sin\xi t'}{\xi}\\
 & +2\lim_{a\rightarrow0}\int_{a}^{\infty}d\xi\xi\sin\xi tG(\xi)\left(-\frac{\pi}{2}\cos\xi t'\right).
\end{eqnarray}
Similar expressions arise in Eq.~(\ref{eq:ThetaMinusSym}) where
the integral over $\xi$ is taken first. Combining all the results
we get, 
\begin{gather}
\Theta_{t+t'}^{(-)}=-\frac{2}{\Delta}\lim_{a\rightarrow0}\int_{a}^{\infty}\frac{d\xi}{\xi}F(\xi)\ln\frac{\xi+a}{\xi-a}\\
+\frac{\pi}{\Delta}\lim_{a\rightarrow0}\int_{a}^{\infty}d\xi F(\xi)\cos\xi t\frac{\sin\xi t'}{\xi}\\
-\frac{\pi}{\Delta}\lim_{a\rightarrow0}\int_{a}^{\infty}d\xi\xi G(\xi)\sin\xi t\cos\xi t'\\
+\frac{\pi}{\Delta}\lim_{a\rightarrow0}\int_{a}^{\infty}d\xi F(\xi)\cos\xi t'\frac{\sin\xi t}{\xi}\\
-\frac{\pi}{\Delta}\lim_{a\rightarrow0}\int_{a}^{\infty}d\xi\xi G(\xi)\sin\xi t'\cos\xi t,
\end{gather}
where we have replaced $\xi'$ with $\xi$ in the last two terms.
Simplifying this expression we get, 
\begin{eqnarray}
 & \Theta_{t+t'}^{(-)}=-\frac{2}{\Delta}\lim_{a\rightarrow0}\int_{a}^{\infty}\frac{d\xi}{\xi}F(\xi)\ln\frac{\xi+a}{\xi-a}\\
 & +\frac{\pi}{\Delta}\lim_{a\rightarrow0}\int_{a}^{\infty}d\xi\left[F(\xi)-\xi^{2}G(\xi)\right]\\
 & \times\frac{\cos\xi t'\sin\xi t+\sin\xi t'\cos\xi t}{\xi}.
\end{eqnarray}
In the first term in the latter, as we take the limit we can take
$F(\xi)\rightarrow F(0)=2\Delta\left[1-2f(\Delta)\right]$ and calculate
the integral, 
\begin{gather}
\int_{1}^{\infty}dx\frac{1}{x}\ln\frac{x+1}{x-1}=\frac{\pi^{2}}{4}.
\end{gather}
In the second term we take into account that, 
\begin{gather}
F(\xi)-\xi^{2}G(\xi)=\frac{2\Delta^{2}}{E}\left[1-2f(E)\right].
\end{gather}
We get, 
\begin{eqnarray}
 & \Theta_{t+t'}^{(-)}=-\pi^{2}\Delta\left[1-2f(\Delta)\right]\\
 & +2\Delta^{2}\pi\lim_{a\rightarrow0}\int_{a}^{\infty}d\xi\frac{\left[1-2f(E)\right]}{E}\frac{\sin\xi(t+t')}{\xi}.
\end{eqnarray}
where the limit operation can be dropped in the last expression. At
$T=0$ this simplifies further to, 
\begin{eqnarray}
 & \Theta_{t+t'}^{(-)}=-\pi^{2}\\
 & +2\Delta\pi\int_{0}^{\infty}d\xi\frac{1}{E}\frac{\sin\xi(t+t')}{\xi}.
\end{eqnarray}
The second term in the last expression can be expressed in terms of
a modified Bessel function, 
\begin{gather}
B_{1}(x)\equiv\int_{0}^{\infty}d\xi\frac{1}{\sqrt{\xi^{2}+\Delta^{2}}}\frac{\sin\xi x}{\xi},
\end{gather}
taking the derivative w.r.t. $x$, 
\begin{gather}
\frac{\partial}{\partial x}B_{1}=\int_{0}^{\infty}d\xi\frac{1}{\sqrt{\xi^{2}+\Delta^{2}}}\cos\xi x=K_{0}(\Delta x).
\end{gather}
Therefore we get, 
\begin{gather}
\Theta_{t+t'}^{(-)}=-\pi^{2}+2\Delta\pi\int_{0}^{t+t'}dxK_{0}(\Delta x)\\
=-2\pi\int_{\Delta(t+t')}^{\infty}dxK_{0}(x).
\end{gather}

Introducing the dimensionless time $\lambda\equiv\Delta t$ we arrive
at the expression in the main text.

\subsection{Second type integral}

We show details of the Fourier integral of the coherence factor shown
in Eq~(\ref{eq:LDOSDiag}), 
\begin{gather}
\Theta_{t}^{(0)}=\frac{1}{\Delta}\int d\xi d\xi'\mathcal{L}_{\xi\xi'}e^{-i\xi t},
\end{gather}
see also Ref.~\onlinecite{BCS}. As above we perform the calculations
using the real time as opposed to the dimensionless time $\lambda$
and introduce the latter at the end of the calculation.

Using the symmetric notations for the coherence factor $\mathcal{L}_{\xi\xi'}$
we write, 
\begin{gather}
\Theta_{t}^{(0)}=\frac{2}{\Delta}\int_{0}^{\infty}d\xi d\xi'\frac{F(\xi)-F(\xi')}{\xi^{2}-\xi'^{2}}\cos\xi t,\label{eq:Thet0Int}
\end{gather}
where the integral is taken in the principal value sense at $\xi=\xi'$, 

\[
\mathcal{P}\int_{0}^{\infty}d\xi d\xi'...=\lim_{a\rightarrow0}\int_{a}^{\infty}d\xi d\xi'...
\]
In the first term in Eq.~(\ref{eq:Thet0Int}) we can integrate right
away, 
\begin{gather}
\mathcal{P}'\int_{0}^{\infty}d\xi d\xi'\frac{F(\xi)}{\xi^{2}-\xi'^{2}}\cos\xi\tau=-\frac{\pi^{2}}{8}F(0).\label{eq:1stInt}
\end{gather}
The second term can be integrated over $\xi$, 
\begin{gather}
\mathcal{P}'\lim_{a\rightarrow0}\int_{a}^{\infty}d\xi d\xi'\frac{F(\xi')}{\xi^{2}-\xi'^{2}}\cos\xi\tau\nonumber \\
=-\lim_{a\rightarrow0}\int_{a}^{\infty}d\xi'F(\xi')\left[\frac{\pi}{2}\frac{\sin\xi'\tau}{\xi'}-\frac{1}{2\xi'}\ln\frac{\xi'+a}{\xi'-a}\right]\nonumber \\
=\frac{\pi^{2}}{8}F(0)-\frac{\pi}{2}\int_{0}^{\infty}d\xi'F(\xi')\frac{\sin\xi'\tau}{\xi'}.\label{eq:2ndInt}
\end{gather}
Substituting Eqs.~(\ref{eq:1stInt},\ref{eq:2ndInt}) into Eq.~(\ref{eq:Thet0Int})
we get, 
\begin{gather}
\Theta_{t}^{(0)}=-\frac{\pi^{2}}{2\Delta}F(0)+\frac{\pi}{\Delta}\int_{0}^{\infty}d\xi'F(\xi')\frac{\sin\xi't}{\xi'}.
\end{gather}

The last integral can be taken explicitly at $T=0$, simplifying, 

\begin{eqnarray*}
 & \Theta_{t}^{(0)}=-\pi^{2}\\
 & +\pi\Delta\int_{0}^{\infty}d\xi\left[\sqrt{\xi^{2}+1}-\xi+\frac{1}{\sqrt{\xi^{2}+1}}\right]\frac{\sin\left(\xi t\Delta\right)}{\xi}.
\end{eqnarray*}
Using modified Bessel functions to represent the integrals\cite{GR},
\begin{gather}
\int_{0}^{\infty}d\xi\frac{\cos\xi x}{\xi+\sqrt{\xi^{2}+1}}=\frac{1-xK_{1}(x)}{x^{2}},\\
\int_{0}^{\infty}dxK_{1}(x)=\frac{\pi}{2},
\end{gather}
we obtain, 
\begin{gather}
\Theta_{t}^{(0)}=-\pi\int_{\Delta t}^{\infty}dx\left[K_{0}(x)+\frac{1-xK_{1}(x)}{x^{2}}\right].
\end{gather}
Introducing dimensionless time $\lambda\equiv\Delta t$ we arrive
at Eq.~(\ref{eq:ThetaZero}). 

The third type of the Fourier integral of the coherence factor arising
in Eq.~(\ref{eq:ThetaPlus}) is obtained in the analogous way.
\end{document}